\documentstyle[11pt,aaspp4,flushrt]{article}
\input psfig
\begin{document}
\title{A Spin 3/5 Black Hole in GRO J1655-40?} 
\author{Andrei Gruzinov}
\affil{Institute for Advanced Study, School of Natural Sciences, Princeton, NJ 08540}

\begin{abstract}
When a bright spot in a high-inclination disk orbits a black hole, its light is periodically lensed and Doppler shifted. If the spots occupy a narrow range of radii, a quasi-periodic oscillation (QPO) will result. I calculate the QPO frequency assuming that the spots appear near $r_{\rm max}$ -- the radius of the maximal proper radiation flux from the disk. The calculated frequency depends on the black hole mass and spin. For the microquasar GRO J1655-40, the black hole mass is known and a 300 Hz QPO was observed. The inferred black hole spin is about 60\% of the maximal. The orbit precession frequency at $r_{\rm max}$ is 20Hz, this QPO was also observed in GRO J1655-40. An approximate formula for $r_{\rm max}$ is given. 
\end{abstract}


\section{Introduction}
GRO J1655-40 is a black hole binary with a $7M_{\odot }$  black hole and a $70^{\circ }$ inclination (Orosz \& Bailyn 1997, but see also van der Hooft et al 1998 and Shahbaz et al 1999). Interpreting the X-ray spectra, Zhang, Cui, \& Chen (1997) suggested that the black hole spin is $s=0.9$ (in units of the maximal spin for a given mass of the hole), while Sobczak et al (1999) give an upper limit $s<0.7$.

Here I interpret the 300 Hz QPO (Remillard et al 1999) as the orbital frequency at the radius of the maximal luminous flux, $r_{\rm max}$, and find $s=0.6$. The orbit precession frequency at $r_{\rm max}$ is 20Hz, this QPO was also observed in GRO J1655-40. 

The QPO model is described in \S 2, where we also give a formula for $r_{\rm max}$. This formula is used to calculate the black hole spin in GRO J1655-40. We discuss an obvious weak point of our model in \S 3, and discuss the associated uncertainty in the calculated spin. We suggest a way to test our model and to improve the accuracy of the black hole spin measurements.

\section{The QPO model}

At high inclinations, a bright spot in the disk becomes brighter when its orbit takes it behind the black hole (Bao \& Ostgaard 1995, Karas 1999) \footnote{In an edge-on disk, a flat spot has zero projected area when it is in front of the hole and a finite projected area when it is behind the hole.}. The light from the spot is also Doppler shifted when the spot orbits the hole. It is natural to expect the orbital frequencies of the spots to be observed as the QPO frequencies, but at what radii should the spots appear? We will consider the following possibility -- the spots mostly appear where the luminous flux from the disk is maximal, near $r_{\rm max}$. 

\begin{figure}[htb]
\psfig{figure=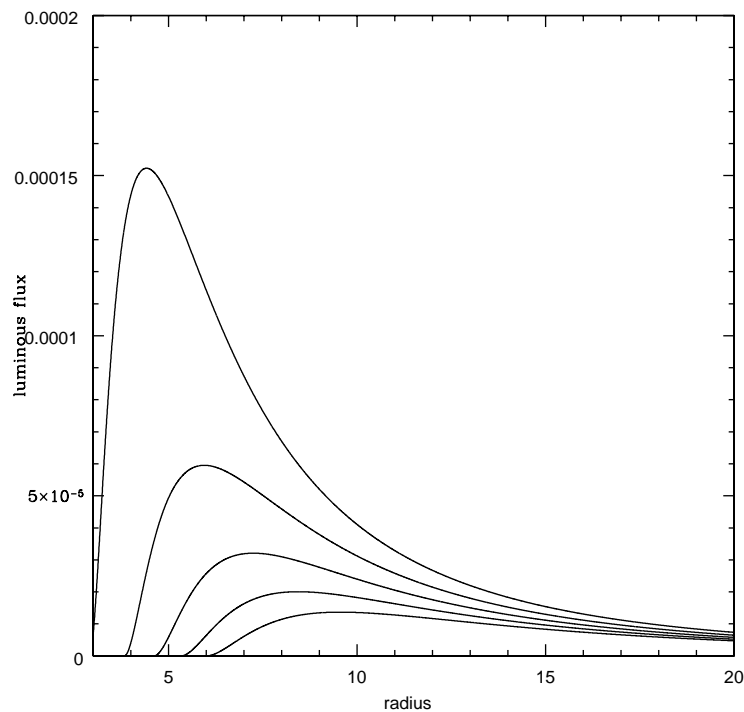,width=4in}
\caption{Luminous flux in units of $\dot{M}$ as a function of radius for spins $s=$ 0., 0.2, 0.4, 0.6, 0.8. Higher curves correspond to higher spins.}
\end{figure}

Page \& Thorne (1974) give an expression for the time-averaged flux of radiant energy (energy per unit proper time per unit proper area) flowing out of the disk, as measured by an observer who orbits with the disk. We plot this flux as a function of radius in Fig. 1 (in Boyer-Lindquist coordinates, with $G=M=c=1$, $M$ is the mass of the hole). It is seen that the flux is relatively flat at large radii, and our assumption -- that the spots mostly appear near $r_{\rm max}$ -- may seem somewhat arbitrary. However, lensing and Doppler shift decrease with radius, and this makes our suggestion more plausible. At any rate, we will quantify the uncertainty in \S 3, assuming that the spots are produced at half the maximal flux \footnote{If the spots are produced in a broad range of radii, the QPO will not be seen at all.}.

We calculated $r_{\rm max}$ numerically. It is found, that 
\begin{equation}
r_{\rm max}=9.55-5.6s-1.78s^{4.3},
\end{equation}
to 1\% accuracy, for $s<0.95$. Equation (1) gives  $r_{\rm max}$ in units of $GM/c^2=1.47{\sl m}$ km, where ${\sl m}=M/M_{\odot }$. The orbital frequency at $r$ is
\begin{equation}
f={1\over r^{3/2}+s},
\end{equation}
in units of $c^3/(2\pi GM)=33{\sl m}^{-1}$ kHz. For a $7M_{\odot }$  black hole, this frequency at $r_{\rm max}$ is equal to 300 Hz if $s=0.58$. 

The orbit of radius $r$ in the Kerr field of spin $s$ has a precession frequency 
\begin{equation}
f_{p}={1\over r^{3/2}+s}-{(1-2r^{-1}+s^2r^{-2})\sqrt{L^2+(1-E^2)s^2}\over (r^2+s^2+2s^2r^{-1})E-2sr^{-1}L}
\end{equation}
where $E$ and $L$ are specific energy and angular momentum of a circular orbit (given in the same paper by Page \& Thorne (1974)). For a $7M_{\odot }$  black hole of spin 0.58, this frequency at $r_{\rm max}$ is equal to 20 Hz. This QPO was also seen in GRO J1655-40. 

Bright spots in the disk are not needed for the observability of the orbit precession as a QPO. Therefore one might expect that orbit precession frequencies corresponding to different radii, both smaller and greater than $r_{\rm max}$, can be observed as QPOs. Indeed QPO frequencies between 10 and 30 Hz have been seen. 

\section{Uncertainties, Tests}

Consider the uncertainty in the calculated spin of the hole in GRO J1655-40. Assume that the bright spots actually appear father from the hole, at $r_{\rm max2}$, where the luminous flux drops to half the maximum. Then the calculated spin is $s=0.9$ and the precession frequency at $r_{\rm max2}$ is 30 Hz. This frequency was indeed seen, but with this assumption it is unclear why the higher orbit precession frequencies corresponding to somewhat smaller radii (where the luminous flux is higher than at $r_{\rm max2}$) are not observed. 

Now assume that the spots appear closer to the hole, at $r_{\rm max1}$, where the flux again drops to half the maximum. Then the calculated spin is $s=0.2$ and the precession frequency at $r_{\rm max1}$ is 8 Hz. This frequency was also seen, but it is unclear why the lower orbit precession frequencies corresponding to somewhat larger radii (where the luminous flux is higher than at $r_{\rm max1}$) are not observed. 

To summarize, our QPO interpretation and the black hole spin estimate for GRO J1655-40 are consistent with the limited observational data available. It might be possible to test the bright spot hypothesis. Theoretical lightcurves in the Kerr metric can be computed. With less than one photon per orbit, and with finite sizes of the bright spots we will see no caustics. But it might be possible to maximize the calculated QPO amplitude using template lightcurves instead of Fourier harmonics as has been done in the original analysis (Remillard et al 1999). The optimal template will give both the radius of the orbit and the black hole spin.

\acknowledgements I thank John Bahcall and Eliot Quataert for useful discussions. I thank Erik Kuulkers and Vladimir Karas for useful information. This work was supported by NSF PHY-9513835.

\end{document}